\input{aipcheck}

\documentclass[
    ,final            
  ]
  {aipproc}

\layoutstyle{6x9}

\begin{document}

\title{Long-term developments in Her X-1: Correlation between the 
histories of the 35\,day turn-on cycle and the 1.24\,sec pulse period}


\author{R. Staubert}{
  address={Institut f\"ur Astronomie und Astrophysik --
Astronomie, University of T\"ubingen, Germany}
}
\author{S. Schandl}{
  address={Institut f\"ur Astronomie und Astrophysik --
Astronomie, University of T\"ubingen, Germany}
}
\author{D. Klochkov}{
  address={Institut f\"ur Astronomie und Astrophysik --
Astronomie, University of T\"ubingen, Germany}
}

\author{J Wilms}{
  address={Department of Physics, Univ. of Warwick, UK}
}

\author{K. Postnov}{
  address={Sternberg Astronomical Institute, Lomonossov University,
Moscow, Russia}
}
\author{N.~Shakura}{
  address={Sternberg Astronomical Institute, Lomonossov University,
Moscow, Russia}
}

\begin{abstract}
 
We have studied the long-term (1971--2005) behaviour of the 1.24\,sec
pulse period and the 35\,day precession period of Her X-1 and show
that both periods vary in a highly correlated way
(see also Staubert et al. 1997 and 2000). When the spin-up rate decreases,
the 35 day turn-on period shortens. This
correlation is most evident on long time scales ($\sim$2000\,days),
e.g., around four extended spin-down episodes, but also on shorter
time scales (a few 100\,days) on which quasi-periodic variations are
apparent. We argue that the likely common cause is variations of
the mass accretion rate onto the neutron star. 
The data since 1991 allow a continuous
sampling and indicate a lag between the
turn-on behaviour and the spin behaviour, in the sense that
changes are first seen in the spin, about one cycle later in the
turn-on. Both the coronal wind model (Schandl \& Meyer 1994) as well 
as the stream-disk model (Shakura et al.\ 1999) predict this kind of 
behaviour.

\end{abstract}

\maketitle


\section{Introduction}
The LMXB Her X-1/HZ Her shows periodicities on very different
time scales. The spin period of the neutron star is 1.24 sec, the
orbit lasts 1.7 days and the 35 day cycle corresponds to the
precession of the warped accretion disk in the tidal field of the
companion. The binary system shows a high inclination of more
than 80 deg, and therefore the disk covers the neutron
star for the observer temporarily during the 35 day precession,
resulting in strong variations of the X-ray signal.
The underlying clock is not very accurate (Staubert et al. 1983,
Klochkov et al. 2006), but we connect its
temporal behaviour with changes in the rate of transfer of mass 
and (more importantly) of angular momentum.
This is supported by a strong correlation between the duration of
the precession cycle and changes in the spin-up rate where the
latter is due to variations of the angular momentum transfer rate 
(Ghosh \& Lamb 1979). 

We suggest a physical explanation within the framework of
the coronal wind model (Schandl \& Meyer 1994) and/or the
stream-disk model by Shakura et al. 1999). A small variation in
the mass transfer rate from the companion may be sufficient to alter 
the torque on the NS and the shape and tilt of the warped disk, 
changing the spin- and precession-frequency, respectively.

\begin{figure}
\includegraphics[width=0.8\textwidth,height=.7\textheight]{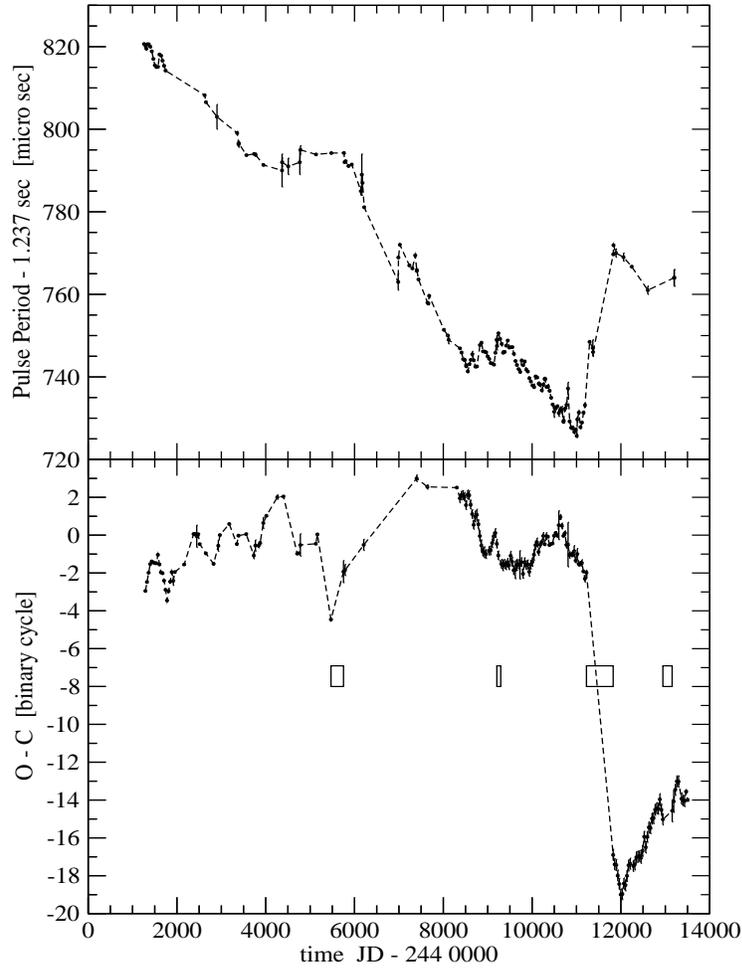}

\vspace{-2mm}
\caption{The 34 year history of the Her X-1 pulse period (upper
panel) and the turn-on data (lower panel), where
(O$-$C) is the difference between observed and calculated turn-on
times. The calculated time follows the epoch of Staubert et al.\
(1983), the 31${^{\rm st}}$ turn-on being on T$_{\mbox {31}}$ = JD
2442410.349 d with a 35\,day model period of 20.5 P$_{\mbox
{orbit}}$ with P$_{\mbox {orbit}}$ = 1.700167788 d (Deeter et al.\
1981). -- The rectangles mark the periods of the four historical Anomalous
Low periods (Parmar et al.\ 1985, Vrtilek et al.\ 1994,
Parmar et al.\ 1999, Coburn et al.\ 2000, Boyd et al.\ 2004, Still \& Boyd\
2004).}
\end{figure}

\section{Data Base}

\subsection{The Spin Period Data}

The historical data for the 1.24\,s X-ray period (the spin
of the NS) are compiled from the original literature. A complete
list will appear in Staubert et al. (2006). 
Since 1991 pulse periods were regularly measured by BATSE onboard of CGRO. 
We have made use of the publically distributed pulsar data as well as lists 
kindly provided by R. Wilson, and data from our own pointed RXTE observations 
(e.g. Stelzer et al.\ 1997, Kuster et al.\ 2005) and the public RXTE archive. 
The pulse period development from its
discovery until today is shown in Fig. 1 (top). The average spin-
up trend dP/dt before day 1150 amounts to $\sim$9 ns/day, but with clear
deviations, including episodes of spin-down. A dramatic spin-down event
happened around day 1150, associated with the third historical anomalous low
period (AL~3).

\subsection{The Turn-on Data}

Her X-1 shows a 35 day flux modulation supposedly due to the occultation of
the NS by a precessing warped accretion disk. Turn-ons, the rise of the
X-ray signal towards the Main-On, are observed since the discovery of Her
X-1 in 1972 (Tananbaum et al.\ 1972). Our historical data set is based on
Staubert et al.\ (1983). Additionally, we determined
turn-ons from the occultation and pulsed flux data of BATSE and from the
RXTE All Sky Monitor. These data were taken from the HEASARC archive at
NASA/GSFC. Details of the determination of the turn-on times and a complete
list of turn-ons will be given in Staubert et al.\ (2006). The
fluctuations of the turn-on times can be expressed by the "(O$-$C)"-diagram
(Fig.~1, lower panel), which shows the difference between the observed
turn-on time and the calculated turn-on time assuming a model 35\,day period
equal to 20.5 P$_{\mbox {orbit}}$ (positive values correspond to turn-ons
which are observed later than the calculated one).

\vspace{-2mm}

\section{Observational Results}

The mean general spin-up of the neutron star ($\sim$9\,nsec/d) is modified
by significant structure: most apparent are four periods of extended
spin-down. These events happen over a time scale of about 2000 days 
and occur near the times of extended periods of low flux 
(so called Anomalous Lows, AL), 
as marked by the rectangles in Fig.~1. In addition, there are
deviations on shorter time scales (a few hundred days) which are of
quasi-periodic nature (on time scales of 400--600 days, see Fig.~3).

The (O$-$C)-diagram is a representation of the history of the 35\,day
period. It appears from Fig.~1 that on the longest time scale the
average period is consistent with 20.5 times the binary period, a
prediction made 22 years ago by Staubert et al.\ (1983) when only
the first increasing leg in the (O$-$C)-diagram was known.
Individual 35\,day cycles (from one turn-on to the next) are
either 20, 20.5 or 21 times the binary period, with a few cycles
with 19.5 times the binary period just before the onset of anomalous
lows and probably several inside the dramatic AL~3 (see below). 

Figs.~1 and 2 show that the large features in the development of
the pulse period and in the (O$-$C) diagram are highly correlated.
The following global features appear simultaneously:
(1) \emph{maxima in pulse period residuals (PPR)}, corresponding to 
(relative) spin-down,
(2) \emph{minima in (O$-$C)} (going into the minimum means a 
short turn-on period),
(3) the \emph{appearance of Anomalous Lows} around these extrema.

These global features repeat about every 5 years, with large differences in
relative strength and duration and with quite some jitter in the
timing. Extrema in (O$-$C) mean that the average 35\,day period changes
its value (it increases coming out of a minimum, and decreases after a
maximum). The most dramatic event is AL~3 in 1999 (around 
JD-2440000=1150). The most likely
average period during AL~3 was about 33.4\,days. Since the start of the 
observations in 1971, some anomalous lows have apparently been missed 
because of poor sampling and/or intrinsic weakness or shortness.

\begin{figure}
\includegraphics[angle=-90,width=1\textwidth]{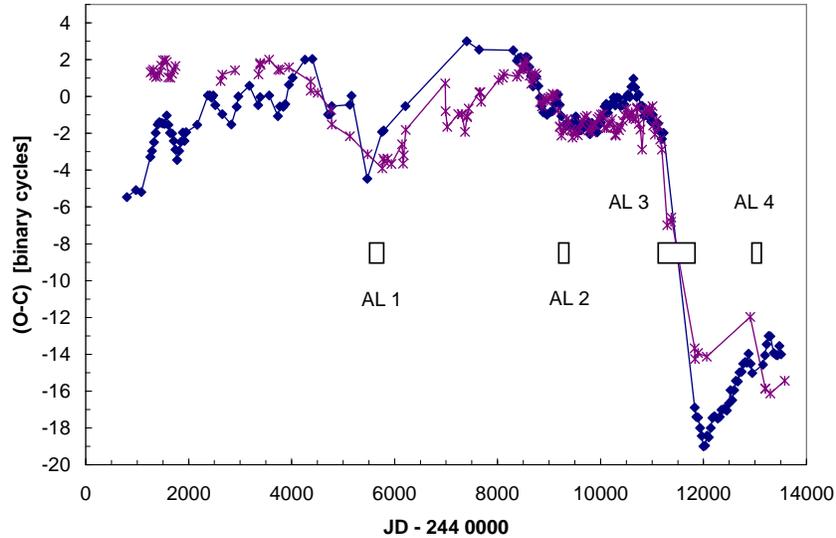}

\vspace*{-2mm}
\caption{The turn-on data (O$-$C) (diamonds) are repeated from Fig.~1. The
second data set (stars) are pulse period residuals (PPR): that is the
separation of the measured pulse periods (upper panel in Fig.~1) from
a linear function representing a constant spin-up. These residuals
were inverted, shifted and scaled in such a way that there is an
optimum fit to the (O$-$C) values for the time frame JD-2440000 between
8500 and 10000. A clear correlation for the global shape of the two
curves is evident. The boxes mark the positions and durations of the
four anomalous lows (AL).}
\end{figure}

\vspace*{2mm}
There are, however, also \emph{clear correlations on shorter time scales} 
(a few hundred days), as shown in Fig.~3. We note
that this kind of correlation has been seen earlier by Bochkarev
et al.\ (1981) and by \"Ogelman et al.\ (1985) on the basis of
smaller sets of data. The data in Fig.~3 are due to the dense 
sampling by BATSE on CGRO. They provide clear evidence for
a time lag between the pulse period evolution and the (O$-$C) evolution,
as already noted by Staubert et al.\ (2000). A formal cross-correlation 
of the pulse period residuals and the (O$-$C) data finds that the
(O$-$C) curve must be shifted to earlier times by $\sim$35\,days to 
reach maximum correlation, meaning that any developments with time are 
seen first in the pulse period and then in the timing of turn-on of
the next 35\,day cycle. 
(We note that the formal lag value of $-35$\,days must not be 
over-interpreted, since the resolution of the (O$-$C) data is, 
by definition, 35\,days. So the likely physical lag is of the order of
a few tens of days).

The changes in pulse period are due to transfer of angular 
momentum to the neutron star. Since the latter should be positively
correlated to the mass accretion rate, one might expect to also see
a correlation between the pulse period residuals and X-ray flux.
While there is some evidence for this (Wilson et al.\ 1997, Staubert et
al.\ 2000, Still \& Boyd\ 2004), the case is not overwhelming and
needs further attention (it will be discussed in 
Staubert et al.\ 2006). Further observational parameters are the 
spectral hardness (Still \& Boyd\ 2004) and the pulse shape 
(Postnov et al.\ 2006).

   
\begin{figure}
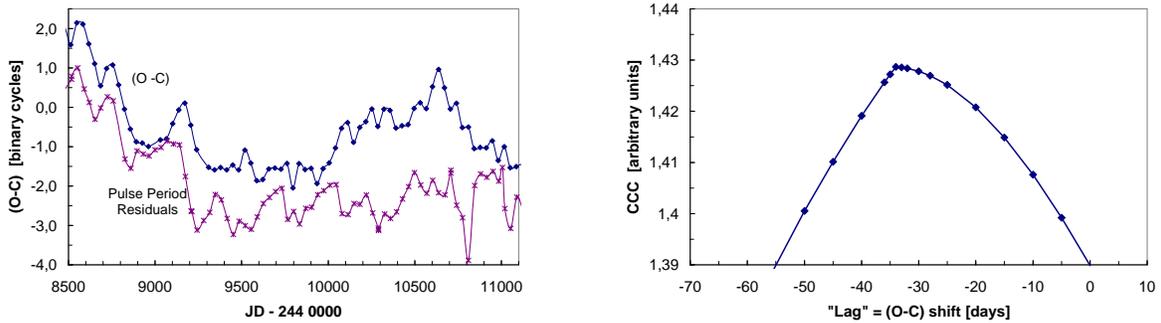

   \includegraphics[angle=-90,width=0.55\textwidth]{Her_Mirax05_fig3.ps} 
   \hfill
   \includegraphics[angle=-90,width=0.55\textwidth]{Her_Mirax05_fig4.ps}
   \caption{Left: (O$-$C) data (upper curve) and pulse period residuals
   (shifted down by 1 unit, lower curve)  for days 8500-11100 (JD-2440000). 
   Right: Cross correlation coefficient (CCC) between the two curves 
   shown left versus an introduced shift of (O$-$C). The
   (O$-$C) lags behind the pulse period residuals by about 35\,days.}
   \label{Fig:plot3&4}
\end{figure}

\vspace*{-3mm}

\section*{Discussion}

A detailed discussion and comparison with models of inclined and
warped accretion disks is not possible here. This will appear
in Staubert et al.\ (2006). Here we note that the observed bahavior
is consistent both with the Coronal Wind model by Schandl \& Meyer
(1994) as well as with the Stream-Disk Interaction model by
Shakura et al. (1999). Under both models we assume that the
motor for the apparent variability is the optical companion which
provides more or less material at the inner Lagrangian point. The
average mass accretion rate of the NS is such that the system
operates close to the equilibrium period with a slight bias
towards spin-up. When the mass accretion rate drops also the spin-up 
rate drops and may even turn to spin-down, in accordance with
standard accretion theory (Ghosh \& Lamb\ 1997). 
The consequences for the observable turn-on times are as observed:
within the framework of the coronal wind model (Schandl \& Meyer\
1994) the reduced X-ray irradiation of the outer
parts of the disk reduces the coronal wind and its torque on the
accretion disk leading to a generally less inclined disk.
This in turn leads to a faster precession of the disk. Anomalous
lows are observed when the disks inclination is very small and
the view onto the NS is blocked. Within the stream-disk model
reduced mass transfer means weaker dynamical action on the disk
by the gas stream leading again to a less inclined disk with 
faster precession. For both models we expect that the NS feels any 
change in mass transfer first and the response of the accretion disk 
is delayed by its viscous time scale (a few tens of days) 
-- as observed.

While the short-term behaviour of the discussed observables in 
Her~X-1 are subject to considerable noise in the system, we
now believe that the global long-term developments are not due to 
a random walk (as proposed by Staubert et al.\ 1983) but rather
due to two physical reasons: (1) the optical companion provides
\emph{quasi-periodic variations in mass transfer} and (2) the NS itself 
- through its possible \emph{free precession} (Ketsaris et al.\ 2000)
- provides a stable internal clock, forcing 
the precession of the disk to stay close to the NS frequency over 
long time scales, despite the variable forces acting on the disk. 
The changing mass transfer rate may be due to changing illumination 
of the optical companion by the X-ray beam. In this sense the apparent 
5~year period could be associated with a limit cycle due to a positive 
feedback in the binary system.




\vspace*{-2mm}
\begin{theacknowledgments}

We acknowledge the support by DFG under grants Sta 173/31 and
436~RUS~113/717/0-1 and by the corresponding RBFR grant 
RFFI-NNIO-03-02-04003 as well as by DLR under grant No.
50~OR~9205. We thank Ljuba Rodina for her valuable contributions
to the analysis of the RXTE data, particularly the 
accurate determination of the pulse periods.

\end{theacknowledgments}



\section*{References}
{\footnotesize

\noindent

Bochkarev N.G., et al., 1981, Sov.Astron.Let. 14(6), 421

Boyd P., Still, M., Corbet, R., 2004, ATEL 307

Coburn W., et al., 2000, ApJ 543, 351

Deeter J.E., Boynton P.E., Pravdo S.H., 1981, ApJ 247, 1003

Ghosh P., Lamb F.K., 1979, ApJ 234, 296

Ketsaris N.A., et al., 2000, Proc. ``Hot points in Astrophysics'',
Dubna, p.192

Klochkov D., et al. 2006, A\&A, submitted


Nagase F., 1989, PASJ 41, 1

\"Ogelmann H., et al., 1985, Sp.Sc.Rev. 40, 347

Parmar A.N., Pietsch W., McKechnie S., et al., 1985, Nat 313, 119

Parmar A.N., et al., 1999, A\&A 350, L5

Postnov K., et al., 2006, in preparation

Schandl S., Meyer F., 1994, A\&A 289, 149

Schandl S., Staubert R., K\"onig 1997, Proc. 4th COMPTON Symp.,
AIP CP 410, 763

Shakura N., et al., 1999, A\&A 348, 917

Staubert R., Bezler M., Kendziorra E., 1983, A\&A 117, 215

Staubert R., Schandl S., Wilms J., 2000, Proc. 5th COMPTON Symp.,
AIP CP 510, 153

Staubert R., et al., 2006, to be submitted

Still M., Boyd P., 2004, ApJ 606, L135

Tananbaum H., Gursky H., Kellog E.M., et al., 1972, ApJ 174, L143

Vrtilek S.D., Mihara T., Primini F.A., et al., 1994, ApJ 436, L9

Wilson R.B., Scott D.M., Finger M.H., 1997, Proc. 4th COMPTON
Symp., CP 410, 739

}

\end{document}